\documentclass{article}
\usepackage[utf8]{inputenc}
\usepackage{authblk}
\usepackage{setspace}
\usepackage[margin=1.25in]{geometry}
\usepackage{graphicx}
\graphicspath{ {./figures/} }
\usepackage{subcaption}
\usepackage{amsmath}
\usepackage{amssymb}
\usepackage{lineno}
\usepackage{hyperref}
\usepackage{xcolor}

\usepackage[
  backend=bibtex,
  style=nejm,
  citestyle=numeric-comp,
  sorting=none
]{biblatex}
\title{Improving the Precision of First-Principles Calculation of Parton Physics from Lattice QCD}

\author[1*]{Yong Zhao}

\affil[1]{Physics Division, Argonne National Laboratory, Lemont, IL 60439, USA}
\affil[*]{Address correspondence to: yong.zhao@anl.gov}

\date{}

\begin{document}

\maketitle

\begin{abstract}

Large Momentum Effective Theory (LaMET) provides a general framework for computing the multi-dimensional partonic structure of the proton from first principles using lattice quantum chromodynamics (QCD). In this effective field theory approach, LaMET predicts parton distributions through a power expansion and perturbative matching of a class of Euclidean observables---quasi-distributions---evaluated at large proton momenta. Recent advances in lattice renormalization, such as the hybrid scheme with leading-renormalon resummation, together with improved matching kernel that incorporates higher-loop corrections and resummations, have enhanced both the perturbative and power accuracy of LaMET, enabling a reliable quantification of theoretical uncertainties. Moreover, the Coulomb-gauge correlator approach further simplifies lattice analyses and improves the precision of transverse-momentum-dependent structures, particularly in the non-perturbative region. State-of-the-art LaMET calculations have already yielded certain parton observables with important phenomenological impact. In addition, the recently proposed kinematically enhanced lattice interpolation operators promise access to unprecedented proton momenta with greatly improved signal-to-noise ratios, which will extend the range of LaMET prediction and further suppress the power corrections. The remaining challenges, such as controlling excited-state contamination in lattice matrix elements and extracting gluonic distributions, are expected to benefit from emerging lattice techniques for ground-state isolation and noise reduction. Thus, lattice QCD studies of parton physics have entered an exciting stage of precision control and systematic improvement, which will have a broader impact for nuclear and particle experiments.

\end{abstract}

Understanding the quark and gluon structure of the proton is a central goal of nuclear and particle physics. It not only provides essential inputs for precision Standard Model predictions at collider experiments, but also reveals the origin of proton's mass and spin and the mechanism of color confinement, which are fundamental to understanding our visible universe. High-energy scattering offers the only direct probe of the proton's inner structure, which can be described by the parton distribution functions (PDFs) in longitudinal momentum space. This picture can also be extended to the transverse momentum and position spaces to include the transverse-momentum-dependent distributions (TMDs) and generalized parton distributions (GPDs), both of which descend from the 5D Wigner distributions. Deciphering these multi-dimensional distributions is a top priority of current and future experiments at Jefferson Lab~\cite{Dudek:2012vr}, the Electron-Ion Collider at Brookhaven National Laboratory~\cite{Accardi:2012qut,AbdulKhalek:2021gbh} and in China~\cite{Anderle:2021wcy}, and CERN~\cite{Adams:2018pwt}.

As intrinsic properties of the proton, parton distributions are non-perturbative objects in quantum chromodynamics (QCD), the fundamental theory of the strong interaction. For more than half a century, they have been extracted from experiments through global fitting with remarkable successes~\cite{Jimenez-Delgado:2014twa,H1:2015ubc,Alekhin:2017kpj,Hou:2019efy,Bailey:2020ooq,NNPDF:2021njg,ATLAS:2021vod}. Nevertheless, challenges remain in certain kinematic regions, spin-dependent effects, and multi-dimensional structures. Lattice QCD is a powerful first-principles tool that can make non-perturbative predictions to complement the experiments. The early efforts were focused on the Mellin moments of PDFs and GPDs as they are time independent and directly calculable on a Euclidean lattice~\cite{Kronfeld:1984zv,Martinelli:1987si}. However, due to increasing statistical noise and power-divergent operator mixings, only the lowest few moments are accessible. Over the past decade, several new approaches have been developed to access higher Mellin moments or the $x$-dependence~\cite{Liu:1993cv,Detmold:2005gg,Detmold:2021uru,Braun:2007wv,Davoudi:2012ya,Ji:2013dva,Chambers:2017dov,Radyushkin:2017cyf,Ma:2017pxb,Shindler:2023xpd}. Unlike other methods that typically rely on phenomenological reconstruction of the PDFs, 
Large Momentum Effective Theory (LaMET)~\cite{Ji:2013dva,Ji:2014gla,Ji:2020ect} enables direct calculations of their $x$-dependence and has been further extended to the GPDs, TMDs, and Wigner distributions.

LaMET is motivated by Feynman’s parton model~\cite{Feynman:1969ej}, which can be viewed as an effective field theory (EFT) of QCD in the infinite momentum limit~\cite{Ji:2020byp,Ji:2022ezo,Ji:2024oka}. The EFT operator defining the PDF can be expanded and matched from a QCD operator under a large Lorentz boost. To be calculable on a Euclidean lattice, the QCD operator is chosen to be time independent, while the Lorentz boost is realized through a large-momentum external state~\cite{Ji:2013dva}. Consequently, the EFT expansion is controlled by the momentum $P^z$.

\begin{figure}[h]
   \centering
   \includegraphics[width=0.5\textwidth]{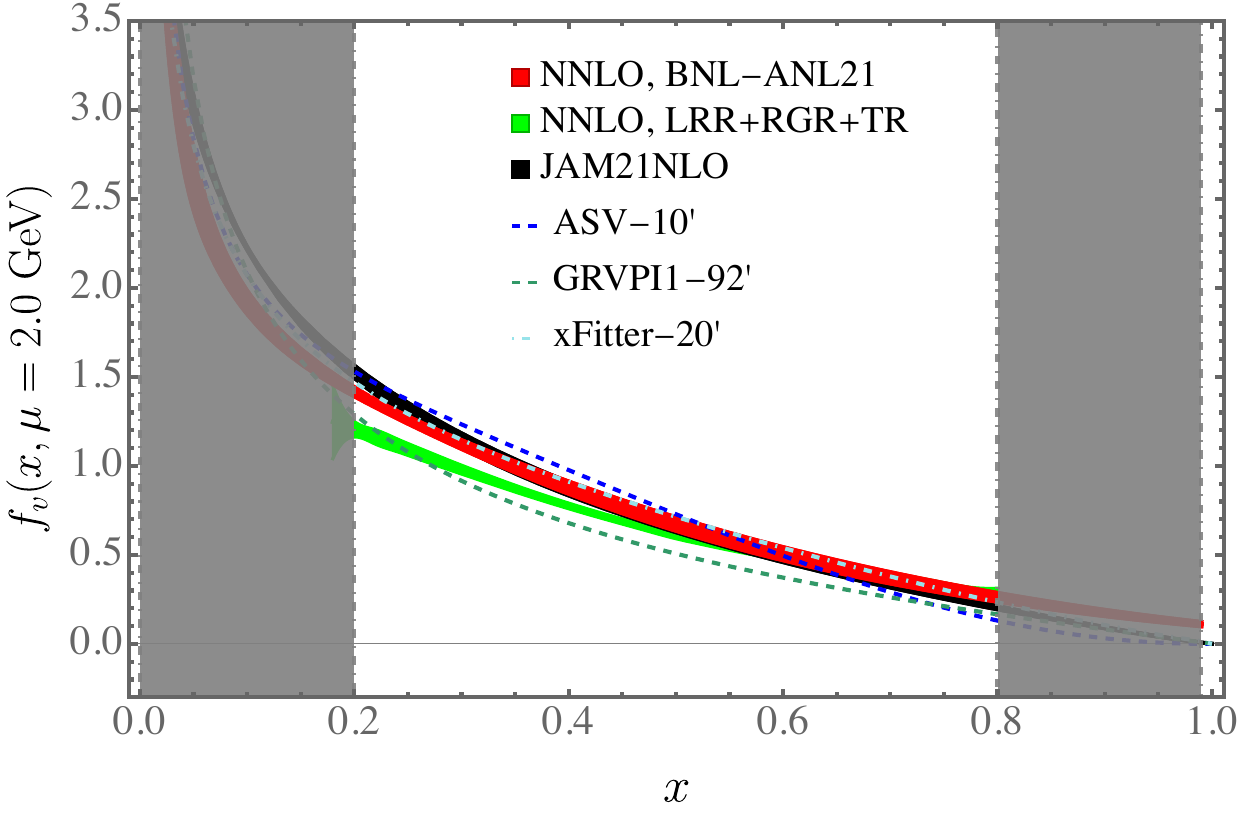}
   \caption{Pion valence PDF from a lattice with spacing $a=0.04$ fm and pion mass $m_\pi=300$ MeV at $P^z=1.94$ GeV~\cite{Gao:2020ito,Gao:2021dbh}. The first NNLO result was obtained by BNL-ANL21~\cite{Gao:2021dbh}, with the updated calculation incorporating LRR, RGR, and TR~\cite{Ji:2024hit}. The lattice results are compared to global fits by GRVP1-92'~\cite{Gluck:1991ey}, ASV-10'~\cite{Aicher:2010cb}, xFitter-20'~\cite{Novikov:2020snp}, and JAM21NLO~\cite{Barry:2021osv}.}
   \label{fig:pdf}
\end{figure}

Since its proposal, LaMET has undergone key developments for systematic uncertainty control. These include proving multiplicative renormalizability for the nonlocal Wilson line operators defining quasi-PDFs~\cite{Ji:2017oey,Ishikawa:2017faj,Green:2017xeu}; establishing non-perturbative renormalization from regularization-independent momentum subtraction~\cite{Constantinou:2017sej,Stewart:2017tvs,Alexandrou:2017huk,Chen:2017mzz} and ratio schemes~\cite{Orginos:2017kos,Braun:2018brg,Li:2020xml,Fan:2020nzz} to the hybrid scheme~\cite{Ji:2020brr}, such as the self-renormalization approach~\cite{LatticePartonLPC:2021gpi}; introducing leading renormalon resummation (LRR)~\cite{Holligan:2023rex,Zhang:2023bxs}; and incorporating higher-order perturbative corrections~\cite{Li:2020xml,Chen:2020ody,Cheng:2024wyu,Ji:2025mvk} and resummations~\cite{Gao:2021hxl,Su:2022fiu,Ji:2023pba,Ji:2024hit}. The hybrid scheme ensures that all renormalized matrix elements can be perturbatively matched to the $\overline{\rm MS}$ scheme, while LRR removes ambiguities in the matching kernel, thereby eliminating the linear power correction in LaMET expansion and improving perturbative convergence. Furthermore, renormalization group and threshold resummations (RGR and TR) systematically resum the large logarithms in the matching kernel as $x\to 0$ and $x\to 1$. The residual power corrections can then be estimated through infinite-momentum extrapolation. Therefore, LaMET enables PDF calculations with controlled theoretical uncertainties for $x \in [x_0, 1-x_0]$ with $x_0\sim \Lambda_{\rm QCD}/P^z$. Figure~\ref{fig:pdf} presents a calculation of the pion valence PDF at $P^z = 1.94$ GeV using next-to-next-to-leading order (NNLO) matching with LRR, RGR, and TR~\cite{Ji:2024hit}. The error band reflects both statistical and scale-variation uncertainties, the latter becoming uncontrolled outside $0.2\lesssim x \lesssim 0.8$ as perturbation theory breaks down. Across this range, the combined uncertainties from statistics, scale variation, higher order and power corrections are estimated to be from 10\% to 20\%~\cite{Gao:2021dbh}. Remarkably, the lattice prediction agrees well with global fits for $0.4<x<0.6$, despite using entirely independent methods. The discrepancies outside this range may arise from residual lattice systematics, including discretization, unphysical quark masses, and finite-volume effects, as well as from uncertainties in the global fits of the pion valence PDF. Therefore, continued efforts are necessary to fully control the systematic uncertainties for a definitive comparison.

The hybrid scheme and LRR can be readily applied to the GPDs and light-cone distribution amplitudes~\cite{Lin:2023gxz,Holligan:2023rex,Holligan:2023jqh,Ding:2024saz}, as they are defined from the same operators as the PDFs. Moreover, the corresponding RGR and TR have been derived~\cite{Baker:2024zcd,Holligan:2025ydm}. Notably, a new method has been developed to extract quasi-GPD matrix elements from the asymmetric frame on the lattice~\cite{Bhattacharya:2022aob}, which significantly reduces the computational cost compared to the symmetric frame.

The TMDs and Wigner distributions are more complex because they involve both a beam and a soft function, and the Collins-Soper (CS) kernel for their evolution can also be non-perturbative. To date, LaMET remains the only approach that can calculate the $x$-dependence of TMDs and Wigner distributions~\cite{Ji:2014hxa,Ji:2018hvs,Ebert:2018gzl,Ebert:2019okf,Ebert:2019tvc,Ji:2019sxk,Ji:2019ewn,Ebert:2020gxr,Ji:2020jeb,Ji:2021znw,Ebert:2022fmh,Deng:2022gzi,Schindler:2022eva} with soft function subtraction~\cite{Ji:2019sxk}, as well as the CS kernel~\cite{Ji:2014hxa,Ebert:2018gzl,Shanahan:2020zxr}. Recently, a state-of-the-art calculation of the quark CS kernel was carried out at physical quark masses with continuum limit and next-to-next-to-leading-logarithmic matching~\cite{Avkhadiev:2023poz,Avkhadiev:2024mgd}, achieving sufficient precision to differentiate among parameterizations in global analyses. The first lattice results for the soft function and proton and pion TMDs are also now available~\cite{LatticeParton:2020uhz,Li:2021wvl,LatticePartonCollaborationLPC:2022myp,LatticeParton:2024mxp}. 

Traditionally, the TMD is calculated from a gauge-invariant correlator with a staple-shaped Wilson line. However, due to the linear divergence in the Wilson line self-energy, the signal-to-noise ratio (SNR) decays exponentially at large transverse separation $b_T$, making it increasingly difficult to access the non-perturbative region. To overcome this bottleneck, new quasi-observables defined in the Coulomb gauge (CG) without Wilson lines have recently been constructed~\cite{Gao:2023lny,Zhao:2023ptv}, which also admit a LaMET expansion~\cite{Zhao:2023ptv}. Free from linear divergences and preserving 3D rotational symmetry~\cite{Gao:2023lny,Zhang:2024omt}, they not only exponentially enhances the SNR at large $b_T$ but also greatly simplifies lattice renormalization. The efficacy of this method has been demonstrated in calculations of the CS kernel~\cite{Bollweg:2024zet}, soft function~\cite{Bollweg:2025iol}, and pion and proton TMDs~\cite{Bollweg:2025iol,Bollweg:2025ecn}. The exponential improvement of the SNR at large distances also reduces Fourier transform uncertainties, particularly in the gluon sector~\cite{Good:2024iur,Good:2025daz}. Therefore, the CG approach has the potential to become a standard framework for calculating TMDs and Wigner distributions, at least for the charge-parity-even sector~\cite{Zhao:2023ptv}, while also complementing the computations of PDFs and GPDs.

Finally, the precision of LaMET calculations is ultimately limited by the maximum attainable momentum. While finer lattices remain desirable, the recently proposed kinematically enhanced hadron interpolators~\cite{Zhang:2025hyo} provide a practical means of reaching unprecedented momenta at current spacings. Besides, as a common source of lattice systematics, the excited-state contamination~\cite{Alexandrou:2020qtt} could be mitigated through interpolator optimization using the generalized eigenvalue problem approach~\cite{Blossier:2009kd} and the recently developed Lanczos method for matrix-element extraction~\cite{Wagman:2024rid,Hackett:2024xnx,Ostmeyer:2024qgu,Chakraborty:2024exj,Hackett:2024nbe,Abbott:2025yhm}.

To summarize, LaMET enables first-principles calculations of parton physics with systematically improvable uncertainties. With ongoing theoretical and computational advances, LaMET is expected to reach $\sim$10\% or better precision in the valence region, thereby exerting an increasingly significant impact on collider experiments.

\section*{Acknowledgments}

We thank R. Zhang for sharing the pion valence PDF data. We also thank A. Avkhadiev, X. Ji, Y. Su, M. Wagman, and R. Zhang for valuable communications. This material is based upon work supported by the U.S. Department of Energy, Office of Science, Office of Nuclear Physics through Contract No.~DE-AC02-06CH11357. The author also acknowledges travel support by the U.S. Department of Energy, Office of Science, Office of Nuclear Physics under the umbrella of the \textit{Quark-Gluon Tomography (QGT) Topical Collaboration} with Award DE-SC0023646.

\printbibliography

\end{document}